# Comparing different solutions for testing resistive defects in low-power SRAMs

N. Mirabella[1,5]   M. Grosso[2]   G. Franchino[3]   S. Rinaudo[1]   I. Deretzis[4]   A. La Magna[4]   M. Sonza Reorda[5]

[1]*AMS R&D STMicroelectronics s.r.l.*, Catania, Italy
[2]*AMS R&D STMicroelectronics s.r.l.,* Torino, Italy
[3]*AMS R&D STMicroelectronics s.r.l.,* Agrate Brianza (MB), Italy
[4]*IMM-CNR*, Catania, Italy
[5]*Dept. of Control and Computer Engineering, Politecnico di Torino,* Torino, Italy

*Abstract*— Low-power SRAM architectures are especially sensitive to many types of defects that may occur during manufacturing. Among these, resistive defects can appear. This paper analyzes some types of such defects that may impair the device functionalities in subtle ways, depending on the defect characteristics, and that may not be directly or easily detectable by traditional test methods, such as March algorithms. We analyze different methods to test such defects and discuss them in terms of complexity and test time.

## I. INTRODUCTION

Systems complexity is constantly rising as well as the demand for devices consuming a lower amount of power (e.g., for portable devices), and this requires particular methods for reducing current consumption inside logic and memories [1].

For example, in the context of the Internet of Things, devices may be required to stand in idle mode as long as they are waiting for scheduled events or environmental changes. Within these periods it is crucial to keep leakage current to a minimum, in particular for what concerns SRAM cells which may be part of large arrays, and for lower technology nodes.

Due to the large area occupied by SRAMs and to their high level of integration, memories are critical from the quality point of view as well. For these reasons, manufacturing test needs to be very accurate, to detect any kind of defects inside the system, and as fast as possible, to contain costs. This situation is worsened by defects that become evident under particular conditions, only, e.g., when the memory changes its status or operation mode. Most of the time this type of defects occurs inside the cell of the memory. They could be due to parasitic capacitance or resistance between the routing. Faulty vias [2][3], defects in the silicon or other kinds of manufacturing imperfections could also cause an unwanted resistive connection inside the cell.

On other hand, defects on low-power structures involve misbehaviors which can hardly be detected by usual March tests. Depending on the resistive-defect value the system undergoes different effects. This paper analyzes the impact of such resistive defects on the behavior of low-power 6T-SRAM cells and evaluates the effectiveness of different test methods. This study specifically considers the effects of such defects when the back-bias technique [4] is employed to reduce leakage.

In brief, our theoretical and experimental analysis provides overall evidence that some types of effects caused by resistive defects in the memory cell can produce different types of misbehaviors inside the system, under certain conditions and resistance value. This article gives an overview of all these defects and how they can be tested, providing useful guidelines to the test engineer in selecting the best test solution(s).

This paper is organized as follows: in section II we introduce low-power memories and some background about memory testing; Section III describes the impact of the analyzed resistive defects on the cell behavior, specifically referring to a 160nm low-power 6T-SRAM exploiting the back-bias technique and evaluating the behavior of each defect when its size changes; in section IV we introduce the possible tests considered for our analysis and in section V we report the results obtained from electrical simulations. Section VI draws some conclusions, helping the test engineer to select the best mix of test solutions, based on the specific constraints in terms of cost (which is mainly impacted by the test duration) and fault detection capabilities.

## II. BACKGROUND AND MOTIVATIONS

### A. Memory testing

Memory testing plays an important role in modern technologies. The design of a memory often requires using the maximum storage density in the minimum area. So, the more the technology evolves, the more the data storage and complexity increase, thus making the appearance of manufacturing defects inside the system more likely.

Previous studies considered the different defects that may affect each cell in a SRAM [5]: in particular, a special attention was given to resistive defects.

### B. Low-power 6T-SRAM structure

The 6T-SRAM cell considered in this paper is depicted in Fig.1. It is made up of two inverters (composed of transistors $M_1$, $M_2$ and $M_3$, $M_4$, respectively) and two pass-transistors $M_5$ and $M_6$. When writing/reading operations are performed, the WL signal is high, letting the *BL* and *BLB* signals to be connected to the *S* and *SB* nodes, respectively, through $M_5$ and $M_6$, thus enabling either reading or writing the cell value.

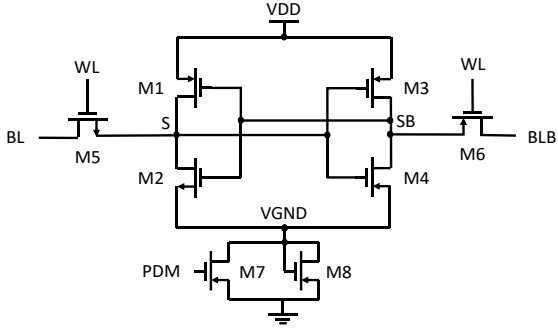

Fig. 1. 6T-SRAM cell with back-bias circuit

If a "0" is written in the cell, the bit lines (*BL* and *BLB*) must be connected to *GND* and *VDD*, respectively. Whereas, when the cell is written with a "1", the *WL* signal activates the pass-transistors $M_5$ and $M_6$, then *BL* and *BLB* are charged to the *VDD* and *GND* value, respectively, thus writing a '1' and a '0' on the *S* and *SB* node, respectively. If no defects occurred inside the cell, the same should stay in the same state as long as a new operation occurs on it.

If a reading operation is performed, the *BL* and *BLB* lines firstly must be pre-charged at *VDD* value. Then, the *S* and *SB* nodes are connected to the *BL* and *BLB* lines, creating a voltage difference detected by a sense amplifier.

The memory under study implements the back-bias technique, a widely used solution that allows the system to reduce the leakage current during the idle periods. This system reduces the rail-to-rail voltage by increasing the voltage of the (virtual) ground node *VGND*. When the control signal PDM (Power Down Mode) is activated, the cell switches from Normal Mode (NM) to Low-Power Mode (LPM), thus activating the back-bias circuit, during which neither writing nor reading operations can be performed in the cell. The back-bias circuitry is usually shared between a set of cells.

*C. Resistive defects and fault models*

The memory cell can be affected by several defects. On a circuit model, some of them can be modeled as Resistive-Bridging [6] and Resistive-Opens defects [7][8][10][10]:

- Resistive-Bridging defects create an unwanted current path between two nodes in the cell which are not intended to be connected.
- Resistive-Open defects increase the resistance of existing paths inside the cell.

Both Resistive-Bridging and Resistive-Open defects may force the cell to misbehave when the corresponding resistance holds specific ranges of values. The functional model of a defect is referred to as *fault*. A wide literature describes the different types of faults that may occur in a SRAM cell [11]. Among them, the following are the most commonly used:

- Stuck-at Fault (SAF), in which the logic value of a cell is always either "0" or "1";
- Transition Fault (TF), when a cell is unable to change its state (0→1 or 1→0) when a write operation is made;
- Data retention Fault (DRF) [12], when a memory cell loses its previously stored logic value after a certain period of time during which it has not been accessed;
- dynamic Data Retention Fault (dDRF) [13], a DRF that occurs when a memory cell loses its previously stored logic value after at least two read or write operations are performed on other cells;
- dynamic Read Destructive Fault (dRDF), that occurs when a write operation immediately followed by a read operation performed on the cell changes the logic state of this cell and returns an incorrect value on the output;
- Read Destructive Fault (RDF), that occurs when a read operation performed on the cell changes the data in the cell itself and returns an incorrect value on the output.

To detect faults in a memory, specific sequences of write and read operations known as March tests are commonly used [14][15]. Often the application of such tests exploits embedded built-in self-test (BIST) logic, to increase test quality and lower costs.

*D. Motivations*

With the continuous decreasing of the MOS-channel size in modern node technologies, circuits are subject to more relevant leakage currents than ever, in particular when the manufactured circuit is low-power.

For this reason, the testing of devices has to be performed in less and less time. To analyze the impact of the different defects and the effectiveness of the different test solutions we used an accurate simulation model of a low-power SRAM cell to evaluate in detail the effect of the different resistive defects possibly affecting it and the ability of the different test methods proposed in literature in detecting them.

III. EFFECTS OF RESISTIVE DEFECTS IN THE LOW-POWER SRAM CELL

This study focuses on some resistive defects that may occur in a low-power SRAM cell. These defects may not only have an effect on the functional behavior of the memory, but also impact on power consumption and/or static-noise-margin. In this paper we will focus on functional effects, only. Indeed, we analyzed the manufacturing defects that can be modelled as resistances within the cell, either as resistive-open or as resistive-bridging defects. When one of such defects occurs, the cell may perform differently from the desired behavior, depending on the defect characteristics. In this section we analyze the impact of each defect.

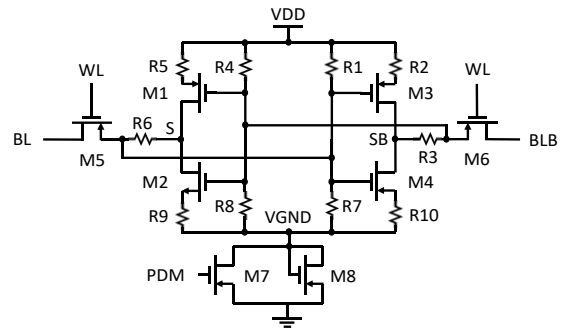

Fig. 2. The resistive defects considered in this study.

Fig. 2 shows the several resistive defects which may arise in the cell; due to its symmetry only three different resistive defects (R1, R2 and R3) have been considered, enough to represent several kinds of manufacturing defects inside a single low-power 6T-SRAM cell.

### A. Resistive path on the cell transistor gate ($R_1$)

The $R_1$ resistance creates a resistive path connection between *VDD* and the gate of $M_3$. This link could produce under particular conditions a failure in the reading and writing operations performed in the cell.

Depending on the $R_1$ value, the typical function of the inverter INV-$M_{3,4}$ may be compromised when the cell is written with a "0". Due to the symmetry of the system, we have the same results with the resistive path $R_8$ between *VGND* and the gate of $M_2$. We have similar effects when the cell is written with a "1" considering INV-$M_{1,2}$ with $R_4$ or $R_7$.

We identified three different resistance ranges in which we have different behaviors of the cell with this defect. In the first range (0÷$R1'$) it does not allow to the cell to be written, whereas the third range (above $R1''$) has no functional effect on the cell no matter whether we are in the NM or in the LPM.

In the middle range (between $R1'$ and $R1''$), the resistive-bridging defect will have no effect in NM, but when the memory moves to LPM and return to NM for a reading function, $R_1$ could cause a change in the data stored in the cell. We define this kind of fault as a Low-Power Retention fault (LPRF), as it behaves as a DRF but it becomes active only when passing through LPM. To summarize, assuming the cell is written with a "0", the behavior of the system with a resistance of an increasing value is the following:

- When the resistance value is very low (in the 0÷$R1'$ range) the cell cannot be written and isolated from the system and it does not work at all, because the *S* and *SB* nodes cannot keep their value. In this case we have a stuck-at fault.
- When the resistance value is in the $R1'$÷$R1''$ range, the cell keeps working well until it passes to LPM and comes back to NM. In that case, the cell cannot preserve its state and switches its value, preserving the faulty value when the cell passes again in the NM state. In this case we have a LPRF.
- When the resistance value is above $R1''$, the cell works correctly in any case because we have a situation very similar to a fault-free SRAM cell.

### B. Resistive open on the cell transistor source terminals ($R_2$)

Another possible resistive defect is $R_2$, consisting in a resistive-open inside the cell. In this case, the behavior of the system is not compromised as long as the resistance value is not high enough to produce a failure. With higher resistance values, this defect can cause misbehaviors in two cases, depending on the resistive defect values:

- When the cell is written and then we quickly perform a read action.
- When the cell is written, then we switch the system to LPM and after switching again to NM we perform a read action.

The effects of the fault can be analyzed referring to four ranges of values for the size of the defect. Assuming the cell is written with a "0", when $R_2$ is present inside the cell, there are the following cases:

- When the resistance is within the range 0÷$R2'$ the system works correctly, even if we perform a read operation either in NM or after a transition between LPM and NM.
- When the resistance is within the range $R2'$÷$R2''$ the cell undergoes a failure when a read after write (RAW) action is performed. This can be modeled as a dRDF.
- When the resistance is above $R2''$, a failure is visible if we perform a RAW in NM (dRDF), or with a read operation right after exiting LPM (LPRF).

Due to the symmetry of the circuit the same behavior can be observed with a resistance that occurs between *VGND* and the source-pin of $M_2$ ($R_9$) or between *VGND* and the source pin of $M_4$ ($R_{10}$) when we write a "1" inside the cell, instead of a "0".

### C. Resistive open between the inverters ($R_3$)

$R_3$ is another type of resistive-open defect that could occur inside the cell. It creates a resistive path between $M_6$ and the *SB* node, thus involving an increased degradation of the voltage value at the *SB* node when the cell is written with a "1" (node *S* at *VDD*, node *SB* at *GND*). Due to the symmetry of the cell, we have the same behavior with $R_6$ when a "0" is written inside the cell. This kind of effect can be modeled as a TF.

The failure of the cell has different ranges of values with respect to the previous defect. Indeed, assuming a "1" written in the cell, there are the following behaviors:

- When the resistance is within the range 0÷$R3'$ the system works correctly, even if we perform a read operation either in the NM or after coming back from the LPM.
- When the resistance is above $R3'$ the cell undergoes a failure when either a read is performed after a write operation regardless of LPM.

## IV. TESTS FOR RESISTIVE DEFECTS IN THE LOW-POWER SRAM CELL

In this Section we summarize our analysis about the ability of different test solutions to detect the considered resistive defects, studying the effects of the different values when their size changes.

### A. $R_1$ case

The faults caused by the $R_1$ defect can be tested through different types of techniques. We analyze the test options in each range case, starting from the minimum to the maximum resistance values.

Due to the misbehaviors caused by $R_1$ (or $R_8$ for symmetry) in the range 0÷$R1'$, any test writing and reading a "0" can detect it. Therefore, a basic March test is sufficient. A similar situation holds for $R_4$ and $R_7$: any test that writes and reads a "1" in the cell can detect the corresponding defect.

For the defects within the $R1'$ ÷ $R1''$ range, a March test is not suitable. However, it is possible to resort to other kinds of tests, such as:

- Low-power retention (LPR) test, corresponding to the following steps. First write into the cell a "0", then enter

LPM, then return to NM, and at last read the cell value. Due to the symmetry of the cell, there are similar behaviors writing a "1" in the cell when considering $R_4$ or $R_7$.

- IDDQ test, corresponding to measuring the quiescent current after writing a "0" (or a "1" for symmetry) in NM. If a defect in this range occurs inside the cell a higher current consumption by the whole system is detectable in NM; indeed we have an higher current consumption when we have already written the cell. Then, after switching to LPM and returning to NM, the current consumption decreases, remaining within the specification limits.

- Read Equivalent Stress (RES) test [17][18][14]. For a particular part of the resistance interval, an alternative methodology can be employed, which does not require current measurements or passages to LPM. This test methodology is based on the Read Equivalent Stress method. The RES test involves repeated reading operations not on the faulty cell but on the other cells in the same row, which cause a stress on the faulty cell. To implement this test solution, we consider a row of cells in the same word line. Firstly, an operation is performed inside the faulty cell, then the other cells are selected. The WL signal continues acting on the other cells but has an impact on the faulty cell because of the stress created by the indirect read action on the same word line. Besides, when the system does not select the cell, the pre-charge circuit stays in the active status and continues charging the bit lines at the *VDD* value. On the other side, when the cell is selected, the pre-charge circuit switches off and an operation can be performed on the cell. It is possible to use a BIST to perform this type of test. Providing that the BIST engine can apply these stimuli, i.e., execute a long enough uninterrupted sequence of selective read operations on the cells of the same row, this method allows an easier to apply and faster test than the tests based on Low-power retention or IDDQ. To extend the effectiveness of the technique, this test can be performed at the minimum *VDD* value admissible by the specifications of the system and the technology.

Above $R1''$ (i.e., when the defect approximates an open circuit) no fault is present.

*B. $R_2$ case*

When considering the fault corresponding to $R_2$ in the range $0 \div R2'$ (where $R2'$ is a relatively small value, depending on the specific cell), there is no functional effect and so no fault to detect.

For $R_2$ and $R_9$ within the $R2' \div R2''$ range, a March RAW (Read After Write) test is sufficient to detect a bit-flip in the cell without passing through LPM.

For $R_2$ and $R_9$ above the $R2''$ value, it is possible to resort to the following types of tests:

- LPR test: write the cell with "0" (or "1" for symmetry), then isolate the cell through *WL* signal by acting on $M_5$ and $M_6$, enter LPM, then return to NM, and at last read the cell value. Due to the symmetry of the cell, there are similar behaviors writing a "1" in the cell and considering $R_5$ and $R_{10}$.

- A March RAW (read after write) test is sufficient to detect a bit-flip in the cell without passing through LPM.

Due to the symmetry of the circuit, we found the same behavior with the other equivalent resistive defects. In particular with $R_9$ when the cell is written with a "0" and $R_5$ or $R_{10}$ when we write a "1" inside the cell.

*C. $R_3$ case*

The fault caused by the $R_3$ defect can be tested through just one type of test. Indeed we have only two ranges considered for the analysis and only for one range a fault is detectable.

When considering the fault corresponding to $R_3$ (or $R_6$, when writing a "1" or a "0" in the cell) in the range $0 \div R3'$, the cell works correctly.

For the $R_3$ (or $R_6$) defect above $R3'$, it is possible to resort to the March RAW (read after write) test that is sufficient to detect a bit-flip in the cell without passing through LPM.

V. EXPERIMENTAL RESULTS

This section presents the experimental results based on the simulation of a 6T SRAM cell in a 160nm STMicroelectronics technology, addressing the defects discussed in the previous sections. For our purpose we used Cadence Virtuoso for the schematics and ELDO simulator for simulations and analysis.

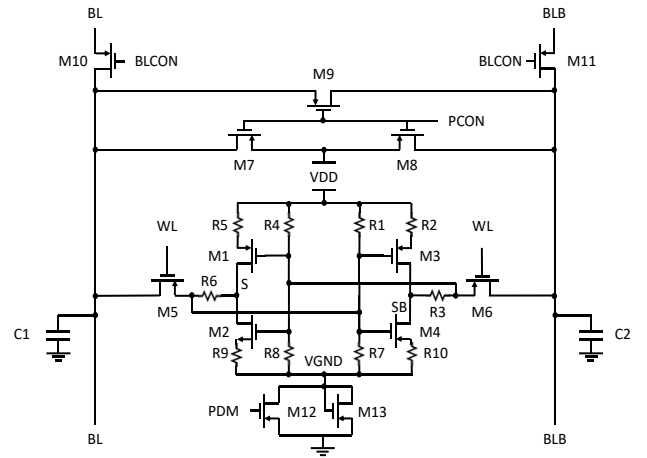

Fig. 3. 6T-SRAM with pre-charge circuit.

Fig. 3 shows a complete low-power SRAM sub-system featuring all the components of a single cell that we used for our simulations. This system is made up of three main components:

- A 6T-SRAM cell with INV-$M_{3,4}$, INV-$M_{1,2}$ and two pass-transistors $M_5$, $M_6$ through which bit-lines can access the cell when the *WL* signal is high.

- A back-bias circuit, increasing the *VGND* value to reduce leakage currents when the cell switches from Normal-Mode (NM) to Low-Power-Mode (LPM).

- A pre-charge circuit ($M_7$, $M_8$, $M_9$), which charges the bit-lines to *VDD* when the cell is not selected for any operation. It is driven by the *PCON* signal that works either before an operation when the cell is selected or when the word line is activated, and other cells are selected for any operation.

To implement our analysis, we properly drive the WL signal in order to write the cell, the PDM signal in order to enable/disable the LPM in the cell, the *PCON* signal to enable/disable the pre-charge circuit, the *BLCON* signal to enable/disable the bit-lines and thus the column of the cells (we consider them with high-impedance end).

The experimental results of our analysis are illustrated in the following sub-sections for each of the considered defects.

*A. $R_1$ case*

Fig. 4a shows the results of the simulation of a LPR test in the *$R_1'\div R_1''$* range, which detects a bit-flip after switching the mode of the cell with $R_1$ inserted in the cell. For sake of comparison, Fig. 4b depicts the behavior of a of a fault-free SRAM cell when the LPR test is applied. Looking at the figure we can see that after exiting LPM the *SB* voltage node sharply goes down causing a failure in the cell behavior and allowing the defect detection. The LPM is usually in the millisecond range at least, but it has been shortened for clarity.

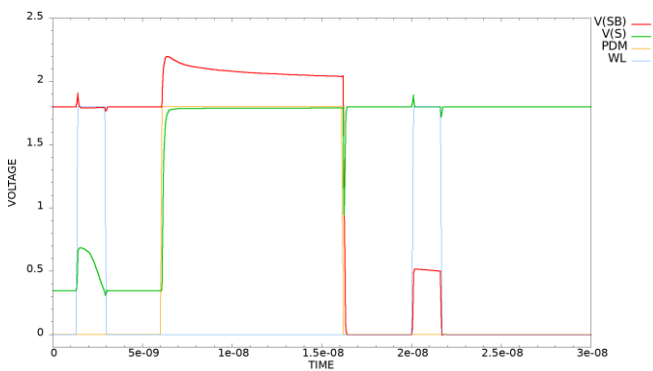

(a)

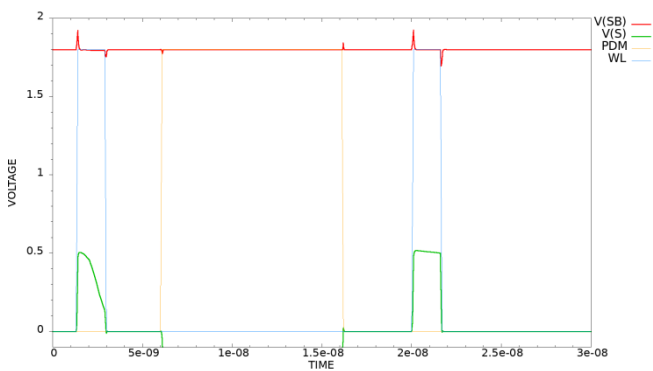

(b)

Fig. 4. (a) Simulation of a LPR test with a bit-flip due to $R_1$; (b) Simulation of a LPR test without defects

In Fig. 5 we consider a resistive value in the sub-range of *$R_1'\div R_1''$*, in which we simulate the effect of the RES test during which we have a bit-flip after several indirect read operations. In this range the considered defect is detected by the RES technique.

According to our simulations, for the considered cells the values for R1' and R1'' are 20kΩ and 35kΩ, respectively. Table I summarizes the defect effects and the fault detection capabilities of each test in each range of values. RES test detects the fault only in a limited sub-range of R1'÷R1'' (around 30kΩ).

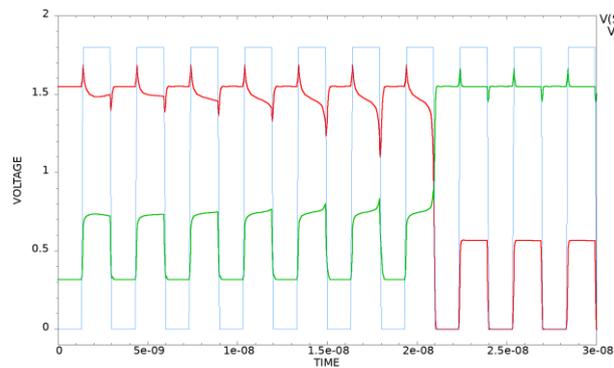

Fig. 5. Simulation of a RES test with a bit-flip due to $R_1$ (allowing the fault detection) after several indirect read operations.

Table 1 – Effectiveness of the different test methods with respect to the $R_1$ defect

|  | <$R_1'$ | $R_1'\div R_1''$ | >$R_1''$ |
|---|---|---|---|
| LPR TEST | Detected | Detected | No Effect |
| March TEST | Detected | Undetected | No Effect |
| IDDQ TEST | Undetected | Detected | No Effect |
| RES TEST | Undetected | Detected (in a sub-range) | No Effect |

*B. $R_2$ case*

Fig. 6 shows the results of the simulation of a LPR test considering the $R_2$ defect above the $R_2''$ value. The simulation detects a bit-flip after reading the data stored when the cell switches back from LPM to NM.

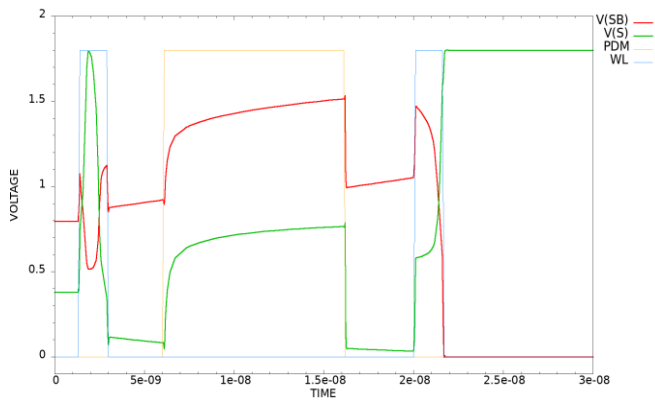

Fig. 6. Simulation of a LPR test with bit-flip because of $R_2$ effect.

According to our simulations, for the considered cells the values for $R_2'$ and $R_2''$ are 3MΩ and 15MΩ, respectively. The effect of every test when the defect size belongs to each range is summarized in the following Table II.

Table II - Effectiveness of the different test methods with respect to the $R_2$ defect

|  | <$R_2'$ | $R_2'\div R_2''$ | >$R_2''$ |
|---|---|---|---|
| LPR TEST | No Effect | Undetected | Detected |
| March TEST | No Effect | Detected | Detected |

## C. $R_3$ case

Fig. 7 shows the results of the simulation of a LPR test considering the $R_3$ defect. The simulation shows that the LPR test detects a bit-flip after reading the stored data when the cell switches back from LPM to NM.

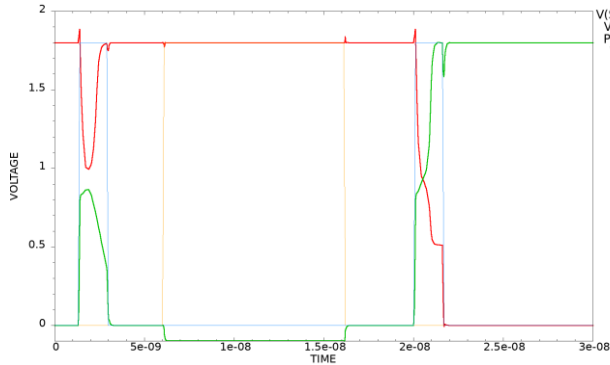

Fig. 7. Simulation of a LPR test with a bit-flip due to $R_3$.

According to our simulations, for the considered cell the values for $R3'$ is 30 kΩ. The effect of every test when the defect size belongs to each range is summarized in the Table III.

Table III – Effectiveness of the different test methods with respect to $R_3$

|  | <R3' | >R3' |
|---|---|---|
| LPR TEST | No Effect | Detected |
| March TEST | No Effect | Detected |

## D. Discussion

To sum up the conclusions of all the analysis that has been performed we may say that traditional retention tests generally require more time than a March test applied by a BIST. Similarly, any test based on the passage through LPM or the IDDQ measure may be relatively long and expensive. Other solutions able to more quickly detect the addressed effects are thus highly welcome. The performed analysis allows to evaluate pros and cons of each case and to choose the best test solution.

We have seen from the data analyzed and the performed tests that for each case the defects that may occur in the circuit could appear in different ways. So, such tests could be relevant to detect such defects.

The best strategy of test must be evaluated case by case considering which technology is used (hence, the likelihood of the different defects, and the range of their values), the quality objectives and which costs we can afford.

## VI. CONCLUSIONS

This paper reports an analysis about the possible defects affecting a 160nm low-power SRAM cell which uses the back-bias technique to reduce the leakage currents. The analysis focuses on three main resistive defects and allows better evaluating the effects of each defect and the effectiveness of different test methods. The analysis can be used by test engineers to select more effectively the test solution(s) to be used for each product. The analysis methodology considering the low-power architecture can be extended also to more advanced technology nodes.

Future works will include:

- the evaluation of the most likely values for such defects through a low-level failure analysis of the specific technology
- the analysis of the non-functional effects induced by the defects
- the assessment of the capability to implement particular BIST tests inside the low-Power SRAM.